 \documentclass[final,5p,times,twocolumn]{elsarticle}

 \usepackage{graphics}
 \usepackage{graphicx}

\usepackage{amssymb}
\usepackage{amsmath}

\journal{Journal of Molecular Spectroscopy}

\begin{document}

\begin{frontmatter}



\title{Submillimeter spectroscopy of H$_2$C$^{17}$O and a revisit of the rotational 
       spectra of H$_2$C$^{18}$O and H$_2$C$^{16}$O}


\author[Koeln]{Holger S.P.~M\"uller\corref{cor1}}
\ead{hspm@ph1.uni-koeln.de}
\cortext[cor1]{Corresponding author.}
\author[Koeln]{Frank Lewen}

\address[Koeln]{I.~Physikalisches Institut, Universit{\"a}t zu K{\"o}ln, 
                Z{\"u}lpicher Str. 77, 50937 K{\"o}ln, Germany}

\begin{abstract}

The rotational spectrum of the formaldehyde isotopologue H$_2$C$^{17}$O was investigated 
between 0.56 and 1.50~THz using a sample of natural isotopic composition. In addition, 
transition frequencies were determined for H$_2$C$^{18}$O and H$_2$C$^{16}$O between 
1.37 and 1.50~THz. The data were combined with critically evaluated literature data 
to derive improved sets of spectroscopic parameters which include $^{17}$O or H nuclear 
hyperfine structure parameters.

\end{abstract}


\end{frontmatter}




\section{Introduction}
\label{introduction}

As a simple four-atomic molecule, formaldehyde, H$_2$CO, also known as methanal, is 
of great fundamental interest. Its rotational spectrum is of great importance for radio 
astronomy. It was only the seventh molecule to be detected in the interstellar medium 
in 1969 \cite{H2CO_det_1969}, the fourth one that was detected by means of radio 
astronomy and only the third poly-atomic molecule; see, for example, the Interstellar 
\& Circumstellar Molecules page\footnote{http://www.astrochymist.org/astrochymist\_ism.html} 
of The Astrochymist\footnote{http://www.astrochymist.org/}. In the detection letter, 
the molecule was observed in absorption toward strong continuum sources, most of them 
dense and warm molecular clouds, so-called hot-cores. Formaldehyde was also detected 
in cold dark clouds, which are also dense, with the $^1$H hyperfine structure (HFS) 
splitting partially resolved \cite{H2CO_dark-clouds_1969}, and in less dense translucent 
\cite{H2CO_translucent_1987} and even less dense diffuse clouds \cite{H2CO_diffuse_1990}. 
It was also detected in the circumstellar envelopes of late-type stars, such as the C-rich 
protoplanetary nebula around V353~Aur \cite{H2CO_C-rich_PPN_1989}, also known as AFGL~618, 
CRL~618, or the Westbrook Nebula, the O-rich protoplanetary nebula around QX~Pup 
\cite{H2CO_O-rich_PPN_1992}, also known as OH231.8+4.2 or the Rotten Egg Nebula, 
or the C-rich asymptotic giant branch star CW~Leo \cite{H2CO_C-rich_AGB_2004}, also 
known as IRC+10216 or the Peanut Nebula. The H$_2$CO molecule was the second molecule 
after OH to be detected in galaxies different from our Milky Way, here the two near-by 
galaxies NGC~253 and NGC~4945 \cite{H2CO_extragal_1974}; it was also detected in more 
distant galaxies \cite{H2CO_B0218_1996}. Formaldehyde is also one of the few molecules 
for which maser activity was not only detected in galactic sources \cite{H2CO_maser_1974}, 
but also in extragalactic sources \cite{H2CO_maser_1986}.

Numerous minor isotopic species were also detected in space, among them H$_2 ^{13}$CO 
\cite{H2C-13-O_1969}, H$_2$C$^{18}$O \cite{H2CO-18_1971}, HDCO \cite{HDCO_1979}, 
and D$_2$CO \cite{D2CO_1990} as the first multiply deuterated molecule in space. 
Unlabeled atoms refer to $^1$H, $^{12}$C, and $^{16}$O. The detection of D$_2$CO 
was made in the Orion~KL region, a site of high-mass star formation, where deuterium 
in formaldehyde was enriched by several orders of magnitudes with respect to the 
interstellar D/H ratio of $\sim 1.5 \times 10^{-5}$. Even higher degrees of 
deuteration were found in the molecular clouds surrounding low-mass proto-stars, 
such as IRAS~16293-2422 \cite{D2CO_IRAS16293_1998}. In fact, deuteration has become 
a means to investigate the evolutionary stage of low-mass proto-stars. The H$_2$CO main 
species may be used to probe the density in denser regions of the interstellar medium 
\cite{H2CO_density_1980} and to determine the kinetic temperature \cite{H2CO_T-kin_1993}. 
The ratios of H$_2 ^{13}$CO to H$_2$C$^{18}$O have been used to infer the 
$^{13}$C$^{16}$O/$^{12}$C$^{18}$O double ratio in molecular clouds 
\cite{H2CO_1316-1218_1981,H2CO_1316-1218_1985}, which in turn may be used to determine 
$^{12}$C/$^{13}$C and/or $^{16}$O/$^{18}$O ratios.

Formaldehyde was also seen in Earth's stratosphere employing microwave limb-sounding with 
the Odin satellite \cite{H2CO_ODIN_2007}; it is more commonly studied in the troposphere 
using infrared or UV/visible spectroscopy among other techniques \cite{H2CO_ODIN_2007}.
The molecule was also detected in the comae of several comets, the first one being comet 
Halley, where it was identified tentatively by infrared spectroscopy 
\cite{H2CO_Halley-IR_1986,H2CO_Halley-IR_1987}, later unambiguously using microwave 
spectroscopy \cite{H2CO_Halley-VLA_1989}.

Formaldehyde was among the first molecules whose rotational spectrum and dipole moment 
were studied by means of microwave spectroscopy \cite{H2CO_rot_dip_1949}. A plethora 
of further studies on the rotational spectrum of H$_2$CO, its isotopologues, not only in 
the ground, but also excited vibrational states were published over the years. 
The rotational spectra of H$_2$CO and its isotopologues began to be explored in the 
terahertz region in the second half of the 1990s, starting with the main isotopologue 
\cite{H2CO_rot_1996}. Investigations of HDCO and D$_2$CO \cite{HDCO_D2CO_rot_1999}, 
H$_2 ^{13}$CO \cite{H2C-13-O_rot_2000}, H$_2$C$^{18}$O \cite{H2CO-18_rot_2000}, and 
again H$_2$CO \cite{H2CO_rot_2003} followed. The most recent study involved studies of 
HDCO and D$_2$CO samples between 1.1 and 1.5~THz \cite{HDCO_D2CO_etc_rot_2015}. 
Important data were obtained for numerous isotopic species with HD$^{13}$C$^{18}$O and 
D$_2 ^{13}$C$^{18}$O being the rarest ones. In addition, the data sets of the already 
well-characterized HDCO and D$_2$CO isotopic species were improved somewhat 
\cite{HDCO_D2CO_etc_rot_2015}. Excited vibrational states of H$_2$CO were also 
investigated up to terahertz frequencies \cite{H2CO_vibs_rot_2009}. 
The spectroscopic parameters were improved by ground state combination differences 
(GSCDs) for H$_2$CO \cite{H2CO_with_GSCDs_2000} and by far-infrared spectra for 
D$_2$CO \cite{D2CO_FIR_2004} and D$_2 ^{13}$CO \cite{D2C-13-O_FIR_2005}. 
In addition, the rotational spectra of formaldehyde in the ground and excited 
vibrational states were used to characterize a spectrometer system based on 
difference frequency generation \cite{H2CO_laser-mixing_2012}.

The most abundant formaldehyde isotopic species, for which terahertz data are lacking, 
is H$_2$C$^{17}$O. Assuming that H$_2 ^{13}$C$^{18}$O is too rare to be detected at 
submillimeter wavelengths, H$_2$C$^{17}$O is the only one for which terahertz data 
are needed. Flygare and Lowe studied five $a$-type $Q$-branch transitions below 14~GHz 
which had $K_a = 1$ and 2 and resolved the $^{17}$O HFS splitting almost completely 
\cite{H2CO-17_rot_1965}. Davies et al. extended the measurements up to 150~GHz with 
HFS resolved to a varying degree \cite{H2CO-17_rot_1980a}. These data were superseded 
by more extensive and more accurate measurements by Cornet et al. which extended up 
to 294~GHz and which were reported only shortly thereafter \cite{H2CO-17_rot_1980b}.

In order to improve the predictions of the rotational spectrum of H$_2$C$^{17}$O especially 
for observations with the Atacama Large Millimeter/submillimeter Array (ALMA) \cite{ALMA_2008}, 
we recorded transitions from 0.56~THz up to 1.50~THz. Additionally, we recorded transitions of 
H$_2$C$^{18}$O and H$_2$C$^{16}$O in the region of 1.37~THz to 1.50~THz. We combined our 
new data with previously reported data for which the initially reported uncertainties were 
critically evaluated. This led to improved spectroscopic parameters which include $^{17}$O 
or H nuclear hyperfine structure parameters.


\section{Experimental details}
\label{exptl_details}

The rotational spectrum of H$_2$C$^{17}$O was recorded in selected regions between 
568 and 658~GHz and between 848 and 927~GHz with the Cologne Terahertz Spectrometer 
(CTS) that is described in detail elsewhere \cite{CTS_1994}. Two phase-locked backward 
wave oscillators (OB~80, OB~82) were used as sources and a magnetically tuned, 
liquid-He-cooled InSb hot electron bolometer (QMC Instruments Ltd.) was used as detector. 
The measurements were carried out in a 4~m long glass cell at room temperature at pressures 
around 1 to 2~Pa. The cell was equipped with windows made from high density polyethylene 
(HDPE). Our study on H$_2$CO \cite{H2CO_rot_2003} may serve as an example for the accuracy 
achievable with the CTS.

Rotational spectra of H$_2$C$^{17}$O, H$_2$C$^{18}$O, and H$_2$C$^{16}$O were recorded 
in selected regions between 1.35 and 1.50~THz using a VDI (Virginia Diodes, Inc.) 
Amplified Multiplier Chain driven by an Agilent E8257D microwave synthesizer as source 
and an InSb bolometer as detector. Measurements were carried out in a 3~m long glass 
cell at room temperature at pressures around 1 to 2~Pa for weaker lines, down to 
around 0.1~Pa for stronger lines. The cell was again equipped with HDPE windows. 
Our study on low-lying vibrational states $\varv_8 \le 2$ of methyl cyanide 
\cite{MeCN_v8_le_2_rot_2015} may serve as an example for the accuracy achievable 
with this spectrometer system.

Formaldehyde was generated by heating a small sample of commercial paraformaldehyde 
briefly with a heat-gun. Frequency modulation was used throughout with demodulation 
at $2f$, which causes an isolated line to appear approximately as a second derivative 
of a Gaussian.


 \begin{figure}
 \begin{center}
  \includegraphics[angle=0,width=7.0cm]{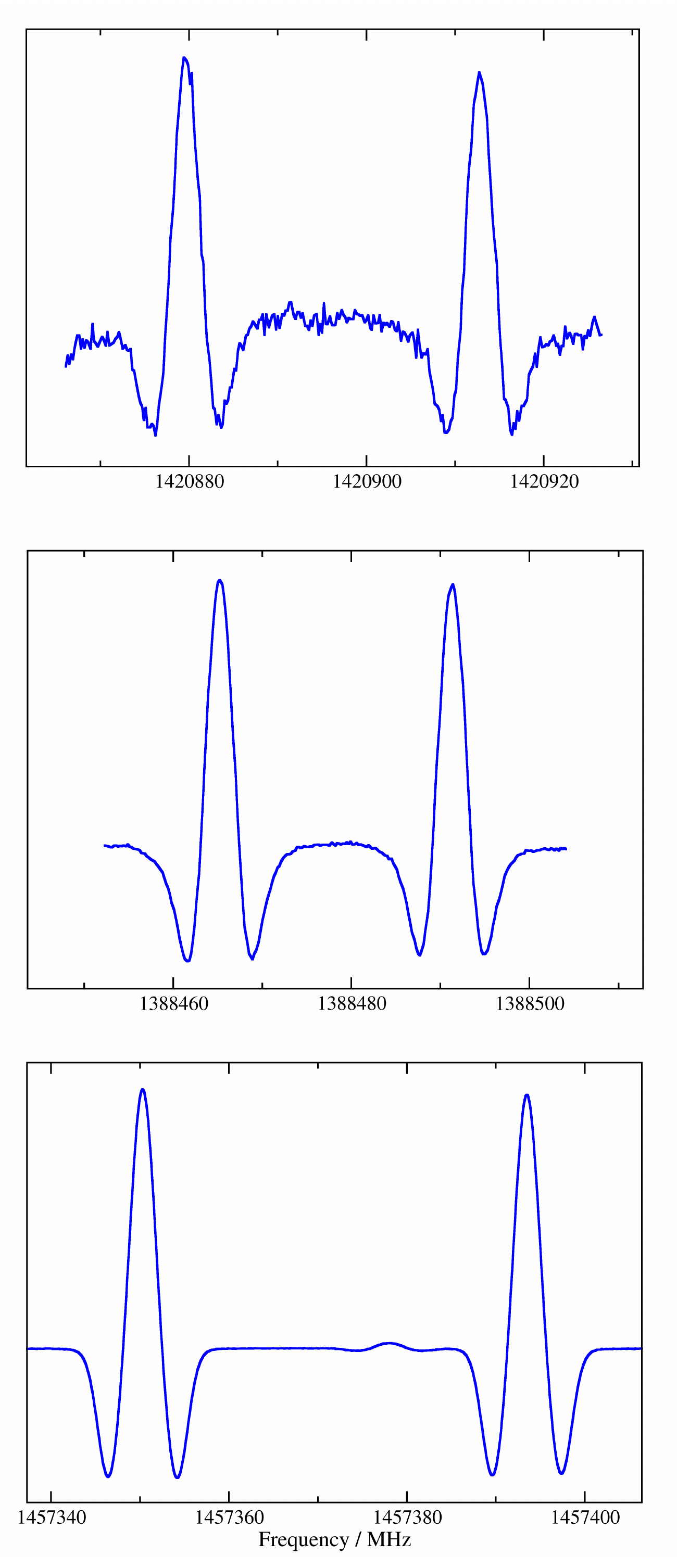}
 \end{center}
  \caption{Detail of the formaldehyde terahertz spectrum displaying the $J = 20 - 19$, 
           $K_a = 5$ rotational transitions of H$_2$C$^{17}$O, H$_2$C$^{18}$O, and 
           H$_2$C$^{16}$O (from top to bottom) with resolved asymmetry splitting. 
           The asymmetry splitting of H$_2$C$^{17}$O is between those of the heavier 
           and the lighter isotopologue, as can be expected. 
           The weak feature near 1457378~MHz in the bottom panel is unassigned.}
  \label{asy-splitting}
 \end{figure}


\section{Spectroscopic analysis}
\label{analysis}

H$_2$C$^{16}$O is an asymmetric top molecule close to the prolate limit ($\kappa = 
-0.9610$ with a dipole moment of 2.3317~D \cite{H2CO-div-isos_dipole_1977} along the 
$a$ inertial axis, which is also the C$_2$ symmetry axis. The asymmetry and the dipole 
moment change only slightly with isotopologue. The two equivalent H nuclei lead to 
spin-statistical weight factors of 1 and 3 for rotational states with $K_a$ even 
(\textit{para}) and odd (\textit{ortho}), respectively. At high resolution, HFS 
splitting may be resolved for \textit{ortho} transitions. This is usually only 
achieved at radio frequencies (RF) or in the microwave (MW) region. The splitting 
was also resolved in astronomical observations of colder environments, but only 
for $K_a = 1$ and low values of $J$.

Even though formaldehyde's proximity to the prolate limit would make Watson's $S$ 
reduction the natural choice for fitting of its rotationally resolved spectra to 
some spectroscopists, use of the $A$ reduction is not so far-fetched. In fact, it 
was the $A$ reduction that was most commonly applied until fairly recently 
\cite{H2CO_rot_1996,HDCO_D2CO_rot_1999}. There was only one detailed consideration 
of the $S$ reduction in earlier reports \cite{H2CO_S-red_1983}, but the results 
present for the main isotopologue were actually slightly worse in the $S$ reduction. 
The advantage of the $S$ reduction has only lately become increasingly apparent 
in most of the diagonal distortion parameters which are smaller in magnitude for 
the sextic distortion parameters in the $S$ reduction compared to the $A$ reduction, 
and the differences are more pronounced in the octic distortion parameters 
\cite{H2C-13-O_rot_2000,H2CO_rot_2003,H2CO_with_GSCDs_2000}. The situation is 
less clear for the off-diagonal distortion parameters, $d_1$, $d_2$, etc. 
in the $S$ reduction, $\delta _K$, $\delta _J$, etc. in the $A$ reduction. 
But it is the large number of off-diagonal distortion parameters needed to fit 
the formaldehyde spectra and their relatively large magnitudes which cause the 
pronounced differences between the two reductions.

Prediction and fitting of the rotational spectra were made with Pickett's SPCAT and 
SPFIT programs \cite{spfit_1991}. Our new data were fit together with previously 
reported line frequencies, and we consulted the original references to check for 
the reported uncertainties. In almost all instances, we used the initially reported 
uncertainties, which is different from some studies where uncertainties had been 
increased considerably, usually without justification. In very few cases of transition 
frequencies with larger residuals, the uncertainties were increased slightly or the 
transition frequencies were omitted from the line lists. Transitions with HFS splitting 
were used as such. In order to keep the line list short, each isotopic species was 
defined twice in its parameter file, with and without HFS. Overlapping HFS or asymmetry 
components were treated in the fit as intensity-weighted averages, in contrast to 
most other fitting programs which treat each overlapping components as one piece 
of information with exactly the assigned frequency, which may increase the rms error 
unless uncertainties were increased beyond the usual extent.

Some higher order parameters were evaluated from other isotopic species, usually 
H$_2$C$^{16}$O, by scaling the parameters with appropriate powers of $B+C$ and $B-C$; 
$A-(B+C)/2$ was very similar among the three species and was not considered for scaling. 
Even if such scaling is not the best choice for all parameters, it is often a good 
approximation. Such scaling was used, for example, for $^{13}$C and $^{15}$N isotopic 
species of methyl cyanide \cite{MeCN_13C-vib_rot_2016}.

There are different sign conventions concerning the nuclear spin-rotation parameters 
and the nuclear spin-nuclear spin coupling parameters. The sign conventions in SPFIT 
are such that in the first case the magnetic moment of H is positive. This convention 
is common nowadays in rotational spectroscopy, e.g., \cite{H2CO-div-isos_dipole_1977}, 
but is opposite to nuclear magnetic resonance and to earlier rotational studies, 
e.g., \cite{H2CO-17_rot_1965,H2CO-17_rot_1980b}. The sign convention in the second 
case is that the nuclear spin-nuclear spin coupling parameters of homo-diatomics are 
negative; there appears to be no clear preference for this or for the opposite sign 
convention.

\subsection{H$_2$C$^{17}$O}
\label{17O-analysis}

The $^{17}$O isotope is the rarest of the stable oxygen isotopes with a terrestrial 
abundance of 0.00038 \cite{composition_elements_2009}. The isotope possesses a nuclear 
spin of 5/2 which gives rise to HFS splitting caused by the nuclear electric quadrupole 
and the nuclear magnetic dipole moments.

Initial predictions of the rotational spectrum of H$_2$C$^{17}$O were generated from 
the data reported by Flygare and Lowe \cite{H2CO-17_rot_1965} and Cornet et al. 
\cite{H2CO-17_rot_1980b}. Both studies resolved $^{17}$O HFS splitting to a different 
degree depending on the quantum numbers and the frequency region; no HFS splitting 
caused by the H nuclei was reported. Initial spectroscopic parameters were taken 
from the latter study which were converted to the $S$ reduction subsequently. 
Additional higher order parameters were derived from H$_2$C$^{16}$O \cite{H2CO_rot_2003}. 
The initially reported uncertainties were used for essentially all transition frequencies 
and essentially all reported HFS information was used. Some modifications were made to 
the list of transition frequencies from Ref.~\cite{H2CO-17_rot_1965}. There was a 
typographical error in the $1_{10} - 1_{11}$ center frequency; an increase by 0.5~MHz 
is agrees almost within uncertainty with the frequency calculated from the final set 
of spectroscopic parameters and was used in the final line list. The remaining data 
were reproduced slightly outside the uncertainties on average. Therefore, uncertainties 
of the more poorly fitting data, $2_{11} - 2_{12}$ center frequency and of two HFS 
splittings of the $6_{24} - 6_{25}$ transition were doubled. In addition, one HFS 
splitting of the $2_{11} - 2_{12}$ transition, involving a weak HFS component, 
was omitted. These modifications affected obviously the partial rms error of this 
data set and, to a lesser extent, the rms error of the entire fit; the parameter 
values and their uncertainties were only slightly affected.

Despite the low $^{17}$O isotopic abundance, the strengths of the formaldehyde absorption 
lines were sufficient to obtain reasonable signal-to-noise ratios for H$_2$C$^{17}$O 
lines in the present study, see Fig.~\ref{asy-splitting}. The detected transitions 
involve $\Delta K_a = 0$ $R$-branch transitions with $7 \le J \le 22$ and $K_a$ up to 7. 
None of the observed transitions displayed HFS splitting, as may be expected.

The spectroscopic parameters determined in the final fit are almost complete up to 
sixth order, only $H_K$ and $H_J$ were kept fixed to the estimated values. In addition, 
two independent quadrupole parameters, $\chi _{aa}$ and $\chi _{bb}$, were determined 
along with three nuclear spin-rotation parameters. $C_{cc}$ was retained in the fit 
because its uncertainty is commensurate with those of $C_{aa}$ and $C_{bb}$. The value 
and the uncertainty of $\chi _{cc}$ were derived from the tracelessness of the quadrupole 
tensor. An edited version of the fit file is available as supplementary material. 
The final spectroscopic parameters of H$_2$C$^{17}$O are given in Table~\ref{parameters} 
together with those of H$_2$C$^{18}$O and H$_2$C$^{16}$O. The rms error of the final fit 
is 0.870, meaning that the experimental data have been reproduced within their 
uncertainties on average. The partial rms errors are 1.019, 0.793, and 0.903 for 
the data from Flygare and Lowe \cite{H2CO-17_rot_1965}, from Cornet et al. 
\cite{H2CO-17_rot_1980b}, and from the present investigation, respectively.


\begin{table*}
\begin{center}
  \caption{Spectroscopic parameters$^a$ (MHz) of formaldehyde isotopologues with 
           $^{17}$O, $^{18}$O, and $^{16}$O along with number of lines and rms error 
           (both unit less).}
  \label{parameters}
{\footnotesize
  \begin{tabular}{lr@{}lr@{}lr@{}l}
  \hline
   Parameter & \multicolumn{2}{c}{H$_2$C$^{17}$O} & \multicolumn{2}{c}{H$_2$C$^{18}$O} 
   & \multicolumn{2}{c}{H$_2$C$^{16}$O} \\
  \hline
$A-(B+C)/2$                   & 246452&.397~(95)     & 247253&.578~(54)     & 245551&.4495~(40)    \\
$(B+C)/2$                     &  35513&.40370~(32)   &  34707&.84108~(25)   &  36419&.11528~(25)   \\
$(B-C)/4$                     &   1148&.454801~(90)  &   1097&.2174152~(59) &   1207&.4358721~(33) \\
$D_K$                         &     19&.448~(33)     &     19&.5203~(151)   &     19&.39136~(53)   \\
$D_{JK}$                      &      1&.257644~(30)  &      1&.2021350~(85) &      1&.3211073~(93) \\
$D_J \times 10^3$             &     67&.10965~(90)   &     64&.30788~(135)  &     70&.32050~(50)   \\
$d_1 \times 10^3$             &   $-$9&.70379~(87)   &   $-$9&.08202~(31)   &  $-$10&.437877~(47)  \\
$d_2 \times 10^3$             &   $-$2&.27013~(64)   &   $-$2&.07709~(38)   &   $-$2&.501496~(33)  \\
$H_K \times 10^3$             &      4&.03           &      4&.03           &      4&.027~(22)     \\
$H_{KJ} \times 10^6$          &      6&.13~(56)      &      2&.615~(77)     &     10&.865~(79)     \\
$H_{JK} \times 10^6$          &      6&.949~(25)     &      6&.380~(9)      &      7&.465~(16)     \\
$H_J \times 10^9$             &      5&.70           &      9&.41~(170)     &      3&.54~(33)      \\
$h_1 \times 10^9$             &     26&.67~(135)     &     27&.23~(51)      &     32&.272~(58)     \\
$h_2 \times 10^9$             &     43&.47~(50)      &     37&.60~(27)      &     47&.942~(74)     \\
$h_3 \times 10^9$             &     13&.87~(27)      &     12&.135~(67)     &     15&.966~(15)     \\
$L_{K} \times 10^6$           &   $-$0&.610          &   $-$0&.610          &   $-$0&.610~(177)    \\
$L_{KKJ} \times 10^9$         &   $-$5&.7            &   $-$5&.5            &   $-$5&.85~(19)      \\
$L_{JK} \times 10^9$          &      0&.35           &      0&.33           &      0&.367~(85)     \\
$L_{JJK} \times 10^9$         &   $-$0&.098          &   $-$0&.091          &   $-$0&.1057~(92)    \\
$l_2 \times 10^{12}$          &   $-$0&.30           &   $-$0&.26           &   $-$0&.345(50)      \\
$l_3 \times 10^{12}$          &   $-$0&.36           &   $-$0&.31           &   $-$0&.427(19)      \\
$l_4 \times 10^{12}$          &   $-$0&.126          &   $-$0&.104          &   $-$0&.1520~(32)    \\
$p_5 \times 10^{18}$          &      2&.60           &      2&.06           &      3&.33           \\
                              &       &              &       &              &       &              \\
\multicolumn{7}{l}{$^{17}$O hyperfine parameters}                                                  \\
  \hline
$\chi _{aa}$                  &   $-1$&.903~(16)     &       &              &       &              \\
$\chi _{bb}$                  &     12&.381~(10)     &       &              &       &              \\
$\chi _{cc}$$^b$              &  $-$10&.478~(10)     &       &              &       &              \\
$C_{aa} \times 10^3$          & $-$366&.4~(25)       &       &              &       &              \\
$C_{bb} \times 10^3$          &  $-$26&.5~(8)        &       &              &       &              \\
$C_{cc} \times 10^3$          &      0&.4~(8)        &       &              &       &              \\
                              &       &              &       &              &       &              \\
\multicolumn{7}{l}{$^1$H hyperfine parameters}                                                     \\
  \hline
$S({\rm HH}) \times 10^3$     &       &              &  $-$17&.933~(98)     &  $-$17&.685~(42)     \\
$C_{\| \ast} \times 10^3$ $^c$ &      &              &   $-$3&.391          &   $-$3&.368~(46)     \\
$C_{\perp} \times 10^3$ $^c$   &      &              &   $-$0&.2481         &   $-$0&.2603~(135)   \\
$C_{-} \times 10^3$ $^c$       &      &              &      1&.0943~(136)   &      1&.1292~(80)    \\
$C_{aa} \times 10^3$ $^b$     &       &              &       &              &   $-$3&.629~(35)     \\
$C_{bb} \times 10^3$ $^b$     &       &              &       &              &      1&.998~(20)     \\
$C_{cc} \times 10^3$ $^b$     &       &              &       &              &   $-$2&.519~(22)     \\
no. of lines$^d$              &    181&              &    147&              &   2043&$^f$          \\
rms error$^e$                 &      0&.870          &      0&.735          &      0&.904          \\
    \hline
  \end{tabular}\\[2pt]
}
\end{center}
$^a$\footnotesize{Watson's $S$ reduction was used in the representation $I^r$. Numbers in parentheses 
     are one standard deviation in units of the least significant figures. Parameter values without 
     uncertainties were estimated and kept fixed in the analyses, see end of general part of 
     section~\ref{analysis}.}\\
$^b$\footnotesize{Derived parameter.}\\
$^c$\footnotesize{$C_{\| \ast} = C_{aa} - (C_{bb} + C_{cc})/2$; $C_{\perp} = (C_{bb} + C_{cc})/2$; 
     $C_{-} = (C_{bb} - C_{cc})/4$.}\\
$^d$\footnotesize{Different pieces of information; i.e., a small number of multiple measurements 
     of, e.g., one transition have been counted separately.}\\
$^e$\footnotesize{Value for the entire fit. Additional details at the end of Sections~\ref{17O-analysis}, 
     \ref{18O-analysis}, and \ref{16O-analysis}.}\\
$^f$\footnotesize{Including 1609 GSCDs.}
\end{table*}

\subsection{H$_2$C$^{18}$O}
\label{18O-analysis}

The $^{18}$O isotope has a terrestrial abundance of 0.0020, more than five times that 
of $^{17}$O \cite{composition_elements_2009}. The abundance difference translates into 
a gain of signal-to-noise or a shorter integration time by a factor of $\sim$30 or 
an appropriate combination, see Fig.~\ref{asy-splitting}. Initial predictions of 
its rotational transitions were taken from the Cologne Database for Molecular 
Spectroscopy, CDMS \cite{CDMS_1,CDMS_2}; these data are based on our previous study of 
H$_2$C$^{18}$O \cite{H2CO-18_rot_2000}. The transitions recorded in the present study 
cover $\Delta K_a = 0$ $R$-branch transitions with $19 \le J \le 22$ and $K_a$ up to 11.

Among the previously published data, resolved HFS splitting was reported for two 
$\Delta K_a = 0$ $Q$-branch transition with $K_a = 1$ and $J = 1$ 
\cite{H2CO-div-isos_6cm_1971} and 2 \cite{H2CO-div-isos_2cm_1972}, respectively. 
This HFS information was used in the present investigation especially to facilitate 
astronomical observations. Initial sets of $^1$H HFS parameters were derived from 
the main isotopic species, see Sect.~\ref{16O-analysis}. Neglecting vibrational effects, 
the spin-spin coupling parameters are expected to be equal, and the spin-rotation 
parameters $C_{gg}$ scale with the respective rotational parameters $B_g$. 
A satisfactory fit was obtained with just the spin-spin coupling parameter $S$(HH) 
and $C_{-} = (C_{bb} - C_{cc})/4)$ released. These are the parameters on which the HFS 
splitting of these transitions depends to first order. No combination of three or even 
four $^1$H HFS parameters yielded a significantly better fit. Moreover, the changes 
from the initial parameters were deemed to be too large for some of the parameters 
if more than two parameters were released in the fits. In case of the $J = 1$ 
transition frequencies, the $F = 0 - 1$ and $F = 2 - 2$ HFS components differ by 
$\sim$0.8~kHz, and the transition frequency published for the latter component 
corresponded much better to the intensity-weighted average of the two components. 
Therefore, we assigned the frequency to the intensity-weighted average in the final fit. 
In case of the $J = 2$ transitions, the $F = 2 - 2$ and $F = 2 - 3$ HFS components are 
close in frequency, and the frequency assigned to the stronger $F = 2 - 2$ component 
differed considerably from the calculated position for this component as well as for 
the intensity-weighted average of the two components. Therefore, this transition 
frequency was omitted from the final fit.

All further rotational data were used as in our previous analysis \cite{H2CO-18_rot_2000}. 
These involve a large body of MW and mmW data from Cornet and Winnewisser 
\cite{H2CO-div-isos_rot_1980} along with three RF transition \cite{H2CO-div-isos_RF_1974} 
and one mmW transition \cite{H2CO_div-isos_vibs_rot_1978}. The set of spectroscopic 
parameters determined for H$_2$C$^{18}$O is almost the same as for H$_2$C$^{17}$O, 
except that $H_J$ was released in the fit of the former. An edited version of the fit 
file is available as supplementary material. The final spectroscopic parameters of 
H$_2$C$^{18}$O are also given in Table~\ref{parameters}. The rms error of the entire 
fit is 0.735, indicative of conservative uncertainties in some data sets. The partial 
rms error of the HFS containing data \cite{H2CO-div-isos_6cm_1971,H2CO-div-isos_2cm_1972}
in the fit is 1.065, that from Refs.~\cite{H2CO-div-isos_rot_1980,H2CO-div-isos_RF_1974} 
are 0.825 and 0.567, respectively. Finally, the rms errors of our previous 
\cite{H2CO-18_rot_2000} and present studies are 0.556 and 0.903, respectively.

\subsection{H$_2$C$^{16}$O}
\label{16O-analysis}

Initial predictions of the rotational transitions of the main isotopic species were 
also taken from the CDMS \cite{CDMS_1,CDMS_2}; these data are based on our previous 
study of H$_2$C$^{16}$O \cite{H2CO_rot_2003}. In the present investigation, frequencies 
were determined for $\Delta K_a = 0$ $R$-branch transitions with $18 \le J \le 21$ and 
$K_a$ up to 15, for four $\Delta K_a = 2$ transitions, and for one $\Delta K_a = 0$ 
$Q$-branch transition with $J = 26$ and $K_a = 1$.

In order to determine the best possible set of HFS parameters, in particular for 
astronomical observations, we evaluated the information in the available original 
reports because effects of HFS were usually omitted in previous studies 
\cite{H2CO_rot_1996,H2CO_rot_2003,H2CO-div-isos_rot_1980,H2CO_div-isos_vibs_rot_1978}. 
In the course of this process, we noticed that uncertainties of previous data were 
increased in Ref.~\cite{H2CO_rot_1996} to usually 1~kHz in cases in which the originally 
reported uncertainties were smaller than this value. The most likely explanation would 
be the difficulty to reproduce the data to within the reported uncertainties. This, in 
turn, may be explained by the reluctance to use a sufficiently large set of off-diagonal 
distortion parameters or by the adherence to the A reduction. In addition, uncertainties 
appeared to have been increased for transitions with unresolved asymmetry splitting for 
which the calculated asymmetry splitting was much larger than the uncertainties.

As in the case of H$_2$C$^{18}$O, resolved HFS splitting was reported for two 
$\Delta K_a = 0$ $Q$-branch transition with $K_a = 1$ and $J = 1$ 
\cite{H2CO-div-isos_6cm_1971} and 2 \cite{H2CO-div-isos_2cm_1972}, respectively; 
the $J = 2$, $F = 2 - 2$ transition frequency omitted for H$_2$C$^{18}$O was also 
omitted for H$_2$C$^{16}$O. Further HFS information originated from an RF investigation 
of H$_2$C$^{16}$O \cite{H2CO_RF_1973}.

Hyperfine free transition frequencies were taken from Ref.~\cite{H2CO_rot_1996} 
with additional original data 
\cite{H2CO-div-isos_rot_1980,H2CO_div-isos_vibs_rot_1978,H2CO_RF_1973,H2CO_RF_1968,H2CO_RF_1977,H2CO_RF_1981,H2CO_review_1972}. 
Further data come from our previous study \cite{H2CO_rot_2003}, from a study of a 
spectrometer system employing difference frequency generation \cite{H2CO_laser-mixing_2012}, 
and from GSCDs generated from IR spectra in the 3.5~$\mu$m region \cite{H2CO_IR-3p5mu_2006}
which were used in a previous ground states study \cite{H2CO_with_GSCDs_2000}.

In almost all instances, we use here the originally reported uncertainties; only in 
very few cases uncertainties were increased slightly if residuals were larger than 
the reported uncertainties and if the partial rms error of a given data set was 
substantially larger than 1.0. If residuals were much larger than the reported 
uncertainties, the corresponding transition frequencies were omitted from the final 
fit. Besides the HFS component mentioned before, this applies to three $K_a = 2$ 
RF transitions \cite{H2CO-div-isos_dipole_1977}. Multiple data with MW accuracy 
were retained in the line list if the uncertainties were similar in magnitude. 
The omitted transitions involve mostly far-infrared laser-sideband data with 
uncertainties around 1~MHz \cite{H2CO_rot_1996}.

The set of rotational and centrifugal distortion parameters is essentially identical 
to that of our previous study \cite{H2CO_rot_2003}; the only difference is the inclusion 
of an estimate of $p_5$ as only parameter that was kept fixed in the fit. This parameter 
was derived from our study on H$_2 ^{13}$CO \cite{H2C-13-O_rot_2000}. In addition, 
the nuclear spin-nuclear spin coupling parameter $S({\rm HH})$ and two sets (in two 
different fits) of three nuclear spin-rotation parameters were determined. An edited 
version of one fit file is available as supplementary material. The final spectroscopic 
parameters of H$_2$C$^{16}$O are also given in Table~\ref{parameters}. The rms error of 
the entire fit is 0.904, this value is dominated by the GSCDs, for which the partial 
rms error is 0.947. Numerous other subsets of the data have rms errors around 0.7, 
the remaining RF data from Tucker et al. \cite{H2CO-div-isos_6cm_1971,H2CO-div-isos_2cm_1972}
are at the upper end (1.006), among the larger subsets, rather low values were obtained 
for the Kiel lines (0.333) \cite{H2CO_rot_1996} and the Cologne lines (0.506) from the 
same study \cite{H2CO_rot_1996}. The rms error of our new lines is 0.726.


\section{Discussion and conclusion}
\label{Discussion}

The rotational and centrifugal distortion parameters of H$_2$C$^{17}$O, which have been 
determined through fitting, compare favorably with those of H$_2$C$^{18}$O and H$_2$C$^{16}$O, 
their values are essentially in all instances between those of the heavier and the lighter 
isotopologue, see Table~\ref{parameters}. The value of $h_1$ appears to be an exception, but its 
uncertainty is large, and an increase by two to three times the uncertainty would bring it to 
the expected value. The H$_2$C$^{18}$O value of $H_J$ is larger than the H$_2$C$^{16}$O value, 
but the uncertainty of the former is quite large. Also, the decrease of $H_{KJ}$ from 
H$_2$C$^{16}$O over H$_2$C$^{17}$O to H$_2$C$^{18}$O is more pronounced than would be expected 
from the scaling mentioned above, but the change in the remaining parameters is rather close 
to what would be expected from such scaling.

The improvement in the distortion parameters of H$_2$C$^{17}$O are quite obvious as the 
$R$-branch transitions were extended from $J = 4 - 3$ near 300~GHz to $J = 22 - 21$ near 
1500~GHz. In addition, $K_a$ extends now to 7, up from previously 4. The improvement is also 
pronounced for H$_2$C$^{18}$O as most of the previous data was limited to below 835~GHz with 
two additional transitions near 1.87~THz. The situation is more complex for H$_2$C$^{16}$O. 
The uncertainties of some parameters changed only slightly, decreased by factors of around 
1.5 to 2 for several others, and even by factors of $\sim$4 for $d_1$ and $L_{KKJ}$.

The present $^{17}$O HFS parameters are slightly better determined than those from the initial 
investigation \cite{H2CO-17_rot_1965} as can be expected because of additional data from 
a later study \cite{H2CO-17_rot_1980b}; uncertainties in the later study are surprisingly 
worse in part than those in the earlier study. The spin-spin coupling parameters $S$(HH) 
may appear quite different among the two isotopic species H$_2$C$^{18}$O and H$_2$C$^{16}$O, 
but the differences are less than two times the combined uncertainties. Because of the 
uncertainties, one should take with a grain of salt that the value calculated from 
the ground state HH distance, derived from $A_0$, is $-$17.907~kHz and thus closer to 
the value of H$_2$C$^{18}$O. Inclusion of higher $K_a$ HFS splitting information in the 
fit improved the uncertainty of $C_{aa}$ by almost a factor of 3 and those of $C_{bb}$ 
and $C_{cc}$ by factors of $\sim$4.

Predictions of the rotational spectra of the three formaldehyde isotopologues will be 
available in the catalog section\footnote{https://cdms.astro.uni-koeln.de/classic/entries/} of the 
CDMS~\cite{CDMS_1,CDMS_2}. Edited fit files are deposited as supplementary material. 
In addition, line, parameter, and fit files, along with other auxiliary files, 
will be available in the spectroscopy data 
section\footnote{https://cdms.astro.uni-koeln.de/classic/predictions/daten/H2CO/} 
of the CDMS.



\section*{Acknowledgements}

We acknowledge support by the Deutsche Forschungsgemeinschaft via the collaborative 
research grant SFB~956 project B3.

\appendix

\section*{Appendix A. Supplementary Material}

Supplementary data associated with this article can be found, in 
the online version, at https://doi.org/10.1016/j.jms.2016.10.004.




\begin{thebibliography}{00}

\bibitem{H2CO_det_1969}  
L.E. Snyder, D. Buhl, B. Zuckerman, P. Palmer, 
Phys. Rev. Lett. 22 (1969) 679$-$681. 

\bibitem{H2CO_dark-clouds_1969}  
P. Palmer, B. Zuckerman, D. Buhl, L.E. Snyder, 
Astrophys. J. 156 (1969) L147$-$L150. 

\bibitem{H2CO_translucent_1987} 
A. Heithausen, U. Mebold, H.W. de Vries, 
Astron. Astrophys. 179 (1987) 263$-$267.

\bibitem{H2CO_diffuse_1990} 
A.G. Nash, 
Astrophys. J. Suppl. Ser. 72 (1990) 303$-$322.

\bibitem{H2CO_C-rich_PPN_1989} 
J. Cernicharo, M. Gu{\'e}lin, J. Penalver, J. Mart{\'{\i}}n-Pintado, R. Mauersberger, 
Astron. Astrophys. 222 (1989) L1$-$L4.

\bibitem{H2CO_O-rich_PPN_1992} 
M. Lindqvist, H. Olofsson, A. Winnberg, L.-{\AA}. Nyman, 
Astron. Astrophys. 263 (1992) 183$-$189. 

\bibitem{H2CO_C-rich_AGB_2004} 
K.E.S. Ford, D.A. Neufeld, P. Schilke, G.J. Melnick, 
Astrophys. J. 614 (2004) 990$-$1006.

\bibitem{H2CO_extragal_1974}  
F.F. Gardner, J.B. Whiteoak,
Nature 247 (1974) 526$-$527. 

\bibitem{H2CO_B0218_1996} 
K.M. Menten, M.J. Reid, 
Astrophys. J. 465 (1996) L99$-$L102. 

\bibitem{H2CO_maser_1974} 
D. Downes, T.L. Wilson,
Astrophys. J. 191 (1974) L77$-$L78. 

\bibitem{H2CO_maser_1986} 
W.A. Baan, R. G{\"u}sten, A.D. Haschik,
Astrophys. J. 305 (1986) 830$-$836. 

\bibitem{H2C-13-O_1969}
P.G. Wannier, A.A. Penzias, R.A. Linke, R.W. Wilson, 
Astrophys. J. 157 (1969) L167$-$L171. 

\bibitem{H2CO-18_1971}  
F.F. Gardner, J.C. Ribes, B.F.C. Cooper, 
Astrophys. Lett. 9 (1971) 181$-$183. 

\bibitem{HDCO_1979}  
W.D. Langer, M.A. Frerking, R.A. Linke, R.W. Wilson, 
Astrophys. J. 232 (1979) L169$-$L173. 

\bibitem{D2CO_1990}  
B.E. Turner, 
Astrophys. J. 362 (1990) L29$-$L33.

\bibitem{D2CO_IRAS16293_1998}   
C. Ceccarelli, A. Castets, L. Loinard, E. Caux, A.G.G.M. Tielens, 
Astron. Astrophys. 338 (1998) L43$-$L46. 


\bibitem{H2CO_density_1980}
A. Wootten, R. Snell, N.J. Evans II,
Astrophys. J. 240 (1980) 532$-$546.

\bibitem{H2CO_T-kin_1993}
J.G. Mangum, A. Wootten,
Astrophys. J. Suppl. Ser. 89 (1993) 123$-$153.

\bibitem{H2CO_1316-1218_1981}
F.F. Gardner, J.B. Whiteoak,
Mon. Not, R. astr. Soc. 194 (1981) 37P$-$41P.

\bibitem{H2CO_1316-1218_1985}
R. G{\"u}sten, C. Henkel, W. Batrla,
Astron. Astrophys. 149 (1985) 195$-$198. 

\bibitem{H2CO_ODIN_2007}
P. Ricaud, D. Alexandre, B. Barret, E. Le Flochmo{\"e}n, E. Motte, 
G. Berthet, F. Lef{\`e}vre, D. Murtagh, 
J. Quant. Spectrosc. Radiat. Transfer 107 (2007) 91$-$104; 
and references therein.

\bibitem{H2CO_Halley-IR_1986}
R.F. Knacke, T.Y. Brooke, R.R. Joyce,
Astrophys. J. 310 (1986) L49$-$L53.

\bibitem{H2CO_Halley-IR_1987}
A.C. Danks, T. Encrenaz, P. Bouchet, T. Le Bertre, A. Chalabaev,
Astron. Astrophys. 184 (1987) 329$-$332. 

\bibitem{H2CO_Halley-VLA_1989}
L.E. Snyder, P. Palmer, I. de Pater,
Astron. J. 97 (1989) 246$-$253.

\bibitem{H2CO_rot_dip_1949}  
J. K. Bragg, A.H. Sharbaugh,
Phys. Rev. 75 (1949) 1774$-$1775. 

\bibitem{H2CO_rot_1996}
R. Bocquet, J. Demaison, L. Poteau, M. Liedtke, S. Belov, K.M.T. Yamamada, 
G. Winnewisser, C. Gerke, J. Gripp, T. K{\"o}hler, 
J. Mol. Spectrosc. 177 (1996) 154$-$159.

\bibitem{HDCO_D2CO_rot_1999}
R. Bocquet, J. Demaison, J. Cosl{\'e}ou, A. Friedrich, L. Margul{\`e}s, 
S. Macholl, H. M{\"a}der, M.M. Beaky, G. Winnewisser, 
J. Mol. Spectrosc. 195 (1999) 345$-$355.

\bibitem{H2C-13-O_rot_2000}
H.S.P. M{\"u}ller, R. Gendriesch, L. Margul{\`e}s, F. Lewen, G. Winnewisser, 
R. Bocquet, J. Demaison, U. W{\"o}tzel, H. M{\"a}der, 
Phys. Chem. Chem. Phys. 2 (2000) 3401$-$3404.

\bibitem{H2CO-18_rot_2000}
H.S.P. M{\"u}ller,  R. Gendriesch, F. Lewen, G. Winnewisser, 
Z. Naturforsch. 55a (2000) 486$-$490.

\bibitem{H2CO_rot_2003}
S. Br{\"u}nken, H.S.P. M{\"u}ller, F. Lewen, G. Winnewisser, 
Phys. Chem. Chem. Phys. 5 (2003) 1515$-$1518.

\bibitem{HDCO_D2CO_etc_rot_2015}
O. Zakharenko, R.A. Motiyenko, L. Margul{\`e}s, T.R. Huet, 
J. Mol. Spectrosc. 317 (2015) 41$-$46.

\bibitem{H2CO_vibs_rot_2009}
L. Margul{\`e}s, A. Perrin, R. Jane\v{c}kov{\'a}, S. Bailleux, 
C.P. Endres, T.F. Giesen, S. Schlemmer, 
Can. J. Phys. 87 (2009) 425$-$435.

\bibitem{H2CO_with_GSCDs_2000}
H.S.P. M{\"u}ller, G. Winnewisser, J. Demaison, A. Perrin, A. Valentin, 
J. Mol. Spectrosc. 200 (2000) 143$-$144.

\bibitem{D2CO_FIR_2004}
J. Lohilahti, V.-M. Horneman, 
J. Mol. Spectrosc. 228 (2004) 1$-$6.

\bibitem{D2C-13-O_FIR_2005}
J. Lohilahti, H. Mattila, V.-M. Horneman, F. Paw{\l}owski, 
J. Mol. Spectrosc. 234 (2005) 279$-$285.

\bibitem{H2CO_laser-mixing_2012}
S. Eliet, A. Cuisset, M. Guinet, F. Hindle, G. Mouret, R. Bocquet, J. Demaison, 
J. Mol. Spectrosc. 279 (2012) 12$-$15.

\bibitem{H2CO-17_rot_1965}
W.H. Flygare, J.T. Lowe
J. Chem. Phys. 43 (1965) 3645$-$3653

\bibitem{H2CO-17_rot_1980a}
D.T. Davies, R.J. Richards, M.C.L. Gerry, 
J. Mol. Spectrosc. 80 (1980) 307$-$319.

\bibitem{H2CO-17_rot_1980b}
R. Cornet, B.M. Landsberg, G. Winnewisser, 
J. Mol. Spectrosc. 82 (1980) 253$-$263.

\bibitem{ALMA_2008}
A. Wootten,
Astrophys. Space Sci. 313 (2008) 9$-$12.

\bibitem{CTS_1994}
G. Winnewisser, A.F. Krupnov, M.Y. Tretyakov, M. Liedtke, F. Lewen, 
A.A. Saleck, R. Schieder, A.P. Shkaev, S.V. Volokhov, 
J. Mol. Spectrosc. 165 (1994) 294$-$300.

\bibitem{MeCN_v8_le_2_rot_2015} 
H.S.P. M{\"u}ller, L.R. Brown, B.J. Drouin, J.C. Pearson, I. Kleiner, 
R.L. Sams, K. Sung, M.H. Ordu, F. Lewen, 
J. Mol. Spectrosc. 312 (2015) 22$-$37.

\bibitem{H2CO-div-isos_dipole_1977}  
B. Fabricant, D. Krieger, J.S. Muenter
J. Chem. Phys. 67 (1977) 1576$-$1586. 

\bibitem{H2CO_S-red_1983}
L. Halonen, A.G. Robiette,
J. Mol. Spectrosc. 101 (1983) 440$-$443.

\bibitem{spfit_1991}
H.M. Pickett,
J. Mol. Spectrosc. 148 (1991) 371$-$377.

\bibitem{MeCN_13C-vib_rot_2016} 
H.S.P. M{\"u}ller, B. J. Drouin, J. C. Pearson, M. H. Ordu, 
N. Wehres, F. Lewen, 
Astron. Astrophys. 586 (2016) A17. 

\bibitem{composition_elements_2009} 
M. Berglund, M.E. Wieser, 
Pure Appl. Chem. 83 (2011) 397$-$.

\bibitem{CDMS_1}
H.S.P. M{\"u}ller, S. Thorwirth, D.A. Roth, G. Winnewisser,
Astron. Astrophys. 370 (2001) L49$-$L52.

\bibitem{CDMS_2}
H.S.P. M{\"u}ller, F. Schl{\"o}der, J. Stutzki, G. Winnewisser,
J. Mol. Struct. 742 (2005) 215$-$227.

\bibitem{H2CO-div-isos_6cm_1971}  
K.D. Tucker, G.R. Tomasevich, P. Thaddeus, 
Astrophys. J., 169 (1971) 429$-$440. 

\bibitem{H2CO-div-isos_2cm_1972}  
K.D. Tucker, G.R. Tomasevich, P. Thaddeus, 
Astrophys. J., 174 (1972) 463$-$466. 

\bibitem{H2CO-div-isos_rot_1980}  
R. Cornet, G. Winnewisser, 
J. Mol. Spectrosc. 80 (1980) 438$-$452.

\bibitem{H2CO-div-isos_RF_1974}  
J.C. Chardon, C. Genty, D. Guichon, N. Sungur, J.G. Th{\'e}obald, 
Rev. Phys. Appl., 9 (1974) 961$-$965. 

\bibitem{H2CO_div-isos_vibs_rot_1978}  
D. Dangoisse, E. Willemot, J. Bellet, 
J. Mol. Spectrosc. 71 (1978) 414$-$429.

\bibitem{H2CO_RF_1973}  
J.C. Chardon, D. Guichon
J. Phys., 34 (1973), 791$-$803. 

\bibitem{H2CO_RF_1968}  
M. Takami,
J. Phys. Soc. Japan, 24 (1968) 372$-$376.

\bibitem{H2CO_RF_1977}  
J.C. Chardon, D. Guichon
J. Phys., 38 (1977) 113$-$120. 

\bibitem{H2CO_RF_1981}  
J.C. Chardon, J.J. Miller, 
Can. J. Phys. 59 (1981) 378$-$386. 

\bibitem{H2CO_review_1972}  
D.R. Johnson, F. Lovas, W.H. Kirchhoff
J. Phys. Chem. Ref. Data, 1 (1972) 1011$-$1045. 

\bibitem{H2CO_IR-3p5mu_2006}  
A. Perrin, A. Valentin, L. Daumont,
J. Mol. Struct. 780$-$781 (2006) 28$-$44.


\end{thebibliography}




\end{document}